\documentstyle{article}

\begin{document}
%\runninghead{Nonsingular String Cosmology
% $\ldots$} {Nonsingular String Cosmology
% $\ldots$}
%\normalsize\textlineskip

\thispagestyle{empty}
\setcounter{page}{1}

%\copyrightheading{}			%{Vol. 0, No. 0 (1993) 000--000}

\vspace*{0.88truein}

%\fpage{1}

\centerline{\LARGE \bf Inhomogeneous String Cosmology Solutions \\}
\centerline{\LARGE \bf with Regular Spacetime Curvature}
\vspace*{0.37truein}

\centerline{\footnotesize LUIS O. PIMENTEL\footnote{E-mail: lopr@xanum.uam.mx}}
\vspace*{0.015truein}
\centerline{\footnotesize\it Departamento de Fisica, Universidad Autonoma Metropolitana-Iztapalapa,}
\baselineskip=10pt
\centerline{\footnotesize\it  A. P. 55-534, CP 09340, Mexico D. F. ,Mexico}

\vspace*{0.225truein}
%\publisher{(received date)}{(revised date)}

\vspace*{0.21truein}

\centerline{\Large \bf  abstract }  
\vspace*{0.15truein}

{In this work cosmological models are considered for the 
low energy string cosmological effective action (tree level) in the 
absence of dilaton potential. A two parametric non-diagonal family of analytic solutions is found. The curvature is non singular, however the string coupling diverges exponentially.}{}{}

%\pacs{Pacs:98.80.Hw, 04.20.Jb, 04.40.-b, 04.50.+h}

\section {Introduction}

The following facts provide the motivation to study inhomogeneous
cosmological models: i)the observable universe is not exactly spatially
homogeneous, ii) there is no evidence that the expansion of the universe 
was regular at very early times. Further motivation comes from the desire 
to avoid postulating special initial conditions. Inhomogeneous cosmological model could also be relevant in the study of galaxies and primordial black
holes formation. In recent times several very interesting exact
inhomogeneous cosmological models with perfect fluids and barotropic
equation of state have been found. The solutions of Wainwright and Goode
\cite{wain}, Feinstein and Senovilla \cite{fei}, Van der Bergh and Skea 
\cite{van}have singularities. Two more recent solutions the one by Senovilla
\cite{S1} and the other by Mars \cite{mars} are remarkable because they do not have big-bang singularity and no other curvature singularity is present.

In 1990, Senovilla {\cite {S1}} found a new perfect-fluid diagonal inhomogeneous cosmological solution without big-bang singularity and 
without any other curvature singularity. The matter contents of that solution was radiation. This solution was shown to be geodetically 
complete and satisfying causality conditions such as global
hyperbolicity {\cite{Sa}}. The singularity-free solution was generalized
in a paper {\cite {RS}} to the case of $G_2$ diagonal cosmologies
and all the different singular behaviors were possible. In a recent 
paper \cite{mars} a new perfect-fluid cosmological solution of
Einstein's equations without big-bang singularity or any other curvature
singularities was found. Neither the energy-momentum tensor nor the Weyl
tensor were singular. The equation of state corresponds to  a stiff 
fluid $p= \rho$, with  positive density and non-vanishing everywhere 
and satisfying global hyperbolicity. This solution is
non-diagonal and it belongs to the class B(i) of  Wainwright for $G_2$ 
cosmologies \cite{W1}. This solution was obtained previously by Letelier \cite{lete}, without noticing the non-singular character of it. 
For some interesting properties of $G_2$ geometries see Ref. \cite{kol}.

At earlier times the fluid model for the matter in the universe might
not be applicable, therefore classical or quantum fields should be
considered for the material content of the models. In particular, it is important to consider the cosmological models of the unifying theories. The purpose of this work is to study inhomogeneous cosmological models
in string cosmology. Recently, several authors have considered the 
case of inhomogeneous cosmological models in the low-energy string cosmological effective action with \cite{gmv1,gmv2,V,B&K,fein}and without 
\cite{giovannini} the presence of the antisymmetric tensor field. In these  
studies the simplification of diagonal inhomogeneous metrics were 
considered, in the present work we consider the non diagonal case.  
The question of the non-singular models in string cosmology has been 
considered by several authors \cite{gmv1,gmv2,giovannini,gas,kalo,eas}.
The first nonsingular inhomogeneous string solution with bounded dilaton
( in contrast with the solutions obtained in this paper, that have unbounded dilaton evolution)  was obtained by 
Gasperini, Maharana and Veneziano \cite{gmv1,gmv2}; they used $O(d,d)$ transformations to "boost away " the singularities.

%%%%%%%%%%%%%%%%%%%%%%%%%%%%%%%%%%%%%%%%%%%%%%%%%%
\section {Field equations in the string frame}
%%%%%%%%%%%%%%%%%%%%%%%%%%%%%%%%%%%%%%%%%%%%%%%%%%

In this paper we consider  inhomogeneous cosmological models in 
the low-energy effective action of string theory, in which no higher order 
correction is taken into account. This theory is obtained assuming that 
only the metric and the dilaton field contribute to the background. 
The effective action in the string frame is \cite{action}:
\begin{equation}
 S = -\frac{1}{\lambda_s^2}\int d^4 x \sqrt{-g} e^{-\phi}  
       \Big[ R + \phi^{;a} \phi_{,a}  \Big],
\end{equation}
where $\lambda_s^2$ is the string scale. As mentioned above, 
we are neglecting any dilaton potential as well as the antisymmetric tensor
 $B_{\mu \nu}$; we are also considering the case of critical superstring 
theory in which the cosmological constant vanishes and six  
internal dimensions are frozen. The gravitational field equation and the equation of motion from the above action are:

\begin{equation}
   R_{ab} 
   = - \phi_{;a;b}\;\;,\;\;\Box \phi  = \phi^{;a} \phi_{;a}.
\end{equation}

For the geometry of the cosmological model we consider the metric with local 
spherical symmetry, assuming that the metric coefficients can be factored as 
products of functions of $t$ and $r$ in the following way,

\begin{equation}
ds^2=a_1(t)\; b_1(r)(-dt^2+dr^2) + a_2(t)\; b_2(r)d\varphi^2 + 
a_3(t)[dz + b_3(r)d\varphi]^2.
\end{equation}
This metric possesses a two-dimensional Abelian group of isometries
 acting on space like surfaces, but with neither
of the Killing vectors being hypersurface-orthogonal, i.e., the metric 
is non-diagonal and it belongs to the class B(i) of  Wainwright for $G_2$ 
cosmologies \cite{W1}. The diagonal case has been considered recently by 
Giovannini \cite{giovannini}. 
The field equations for this particular form of the metric are:

%00 comp
\begin{eqnarray}
& &2\frac{\ddot a_1}{a_1}-2\frac{{\dot a_1}^2}{a_1^2}-\frac{{\dot a_1} 
{\dot a_2}}{a_1 a_2}-\frac{{\dot a_1}{\dot a_3}}{a_1 a_3}+
 2\frac{{\dot a_1}{\dot f}g}{a_1}+2\frac{\ddot a_2}{a_2}-
\frac{{\dot a_2}^2}{a_2^2}
+2\frac{\ddot a_3}{a_3}-\frac{{\dot a_3}^2}{a_3^2}-
2\frac{b_1''}{b_1}+2\frac{{b_1'}^2}{b_1^2}\nonumber \\
& &-\frac{{b_1'}{b_2'}}{b_1 b_2}+2\frac{{b_1'}{g'}f }{b_1}
-4\;{\ddot f}g=0,
\end{eqnarray}

%01 comp
\begin{equation}
\frac{{\dot a_1}{b_2'}}{a_1 b_2}-2  \frac{{\dot a_1}{g'}f}{a_1}+
\frac{{\dot a_2}{b_1'}}{a_2 b_1}-  \frac{{\dot a_2}{b_2'}}{a_2 b_2}
+ \frac{{\dot a_3 }{b_1'}}{a_3 b_1}-2  \frac{{\dot f}{b_1'}g}{b_1}
+4{\dot f}{g'}=0, 
\end{equation}

%11 comp
\begin{eqnarray}
& &2\frac{\ddot a_1}{a_1}-2  \frac{{\dot a_1}^2}{a_1}
+ \frac{{\dot a_1}{\dot a_2}}{a_1 a_2} 
+ \frac{{\dot a_1}{\dot a_3}}{a_1 a_3}-2 \frac{{\dot a_1}{\dot f}g}{a_1}
-2 \frac{b_1''}{b_1} +2 \frac{{b_1'}^2}{b_1^2}+ 
\frac{{b_1'}{b_2'}}{b_1 b_2}-2  \frac{{b_1'}{g'}f}{b_1}\nonumber \\
& &-2\frac{b_2''}{b_2} + \frac{{b_2'}^2}{b_2}
-2 \frac{{b_3'}^2 a_3}{a_2 b_2}+4 g'' f=0,
\end{eqnarray}

%22 comp
\begin{eqnarray}
& &2\frac{\ddot a_2}{a_2}-  \frac{{\dot a_2}^2}{a_2^2}
+ \frac{{\dot a_2}{\dot a_3}}{a_2^2(a_2/a_3+b_3^2/b_2)}
-2 \frac{{\dot a_2}{\dot f}g}{a_2} 
+2\frac{\dot a_3 b_3^2}{a_2 b_2}\nonumber \\
& &-\frac{{\dot a_3}^2 b_3^2}{a_2 a_3 b_2} 
-2\frac{{\dot a_3}{\dot f}b_3 ^2g}{a_2 b_2}-2\frac{b_2''}{b_2}
+\frac{{b_2'}^2}{}b_2^2 
+2\frac{b_2' b_3'a_3 b_3}{a_2 b_2} \nonumber \\
& &+2\frac{b_2'g'f}{b_2}
-4\frac{b_3 a_3 b_3}{a_2 b_2}+2
\frac{{b_3'}^2 a_3}{a_2 b_2(a_3 b_3^2-a_2 b_2)}+4  \frac{b_3' g' a_3b_3 f}{a_2 b_2}=0, 
\end{eqnarray}

%23 comp
\begin{equation}
\frac{{\dot a_2}{\dot a_3}}{a_2 a_3}+2  \frac{\ddot a_3}{a_3}
- \frac{{\dot a_3}^2}{a_3^2} -2 \frac{{\dot a_3}{\dot f}g}{a_3}
+ \frac{{b_3'}{b_3'}}{b_2 b_3}-2  \frac{b_3''}{b_3}
+2\frac{{b_3'}^2a_3}{a_2 b_2}+2  \frac{{b_3'}{g'}f}{b_3}=0,
\end{equation}

% 33comp
\begin{equation}
\frac{{\dot a_2}{\dot a_3}}{a_2 a_3}+2  \frac{\ddot a_3}{a_3}
- \frac{{\dot a_3}^2}{a_3^2}-2 \frac{{\dot a_3}{\dot f}g}{a_3}
+2 \frac{{b_3'}^2a_3}{a_2 b_2}=0 , 
\end{equation}

%KG comp
\begin{equation}
\frac{{\dot a_2}{\dot f}}{a_2 f} +\frac{{\dot a_3}{\dot f}}{a_3 f} 
-\frac{{b_2'}{g'}}{b_2 g}+2 \frac{\ddot f}{f}-2  \frac{{\dot f}^2 g}{f}
-2 \frac{g''}{g} +2 \frac{{g'}^2f}{g}=0. 
\end{equation}
Where I have assumed that $\phi = f(t) g(r)$.

%%%%%%%%%%%%%%%%%%%%%%%%%%%%%%%%%%%%%%%%%%%%
\section{Exact  solutions}
%%%%%%%%%%%%%%%%%%%%%%%%%%%%%%%%%%%%%%%%%%%%

I have found the following 
exact solution to the above field equations,

\begin{eqnarray}
ds^2&=& e^{ht}\left[e^{s r^2} \cosh(2pt) \left ( -dt^2 + dr^2 \right ) 
+ r^2 \cosh(2pt) d\varphi^2 + \frac{\left ( dz + p r^2 d\varphi \right )^2 }
{\cosh(2pt)} \right], \nonumber \\
\phi&=& {h t},
\end{eqnarray}

\noindent where $h,s$ and $p$ are constants satisfying the relation 
$h^2=4(s-p^2)$. The range of variation of the coordinates is
%\begin{eqnarray*}
$-\infty < t,z < \infty , \; 0\leq r < \infty , \;  0 \leq
\varphi \leq 2 \pi$.
%\end{eqnarray*}
This spacetime has a well-defined axis of symmetry at $r=0$ where the assumption of Lorentzian geometry is fulfilled ("elementary flatness "){\cite{KK}}, and therefore the
coordinate $r$ can  to be interpreted as a radial cylindrical coordinate.
Even though the metric is inhomogeneous, the dilaton field is homogeneous.
This situation remind us of the converse situation in Brans-Dicke theory 
where for the homogeneous and isotropic case (Friedman Robertson Walker 
metric), it is possible to find solutions with the scalar field being 
inhomogeneous \cite{MCK}. Also for the case with static spherical symmetry 
it is possible to have a solution where the scalar field is not static 
\cite{nbd}(set $J =1$ in the solution of Ref. \cite{nbd}to see the point). 

We notice here that this metric is conformally related to the non-singular 
one obtained by Mars \cite{mars} in general relativity for a stiff fluid. 

%%%%%%%%%%%%%%%%%%%%%%%%%%%%%%%%%%%%%%%%%%%%%%%
\section{Curvature invariants}
%%%%%%%%%%%%%%%%%%%%%%%%%%%%%%%%%%%%%%%%%%%%%%

Since the conformal factor ( $e^{ht}$) and its derivatives are well behaved, then no curvature singularity appears in the solution of the present work.
Here we compute curvature invariants and observe the absence of singularity in them,

\begin{equation}
I_1=R^2={e^{-2ht-\frac{h^2+4p^2}{2}r^2}} \,{h^4}\,
  {{{\rm Sech}(2\,p\,t)}^2}
\end{equation}
\begin{eqnarray}
I_2=R_{ab}  R^{ab}&=&\frac{e^{-2ht-\frac{h^2+4p^2}{2}r^2}}{16}h^2{\rm sech}^4(2 p t)
 \left [   8\,({h^2} - 4\,{p^2})- (h^2+4p^2)^2r^2 + \right. \nonumber \\
&&\left. \{8(h^2+ 4 p^2)-(h^2+4p^2)^2r^2   \}{\rm cosh(4pt)}
+ 16hp\sinh (4pt)\right ]
\end{eqnarray}

\begin{eqnarray}
I_3&=&R_{abcd}R^{abcd} =\frac{e^{-2ht-\frac{h^2+4p^2}{2}r^2}}{32}{\rm sech}^6(2 p t)
 \left [ 24(h^4-8h^2p^2+280p^2) + \right. \nonumber \\
&& (320\,{p^6}+   112\,{h^2}\,{p^4}- 
    4\,{h^4}{p}^2 -3{h^6}   )r^2 +  \nonumber \\
&& \{32(h^4- 4h^2 p^2-168 p^4)-4(h^6+4h^4p^2-16h^2p^4-64p^6)r^2   \}{\rm cosh(4pt)}\nonumber \\
&&\{8(h^4+8h^2p^2+24p^4)-(h^2+4 p^2)^3r^2\}{\rm cosh(8pt)}\nonumber \\
&&\left. + 32h^3p\sinh (4pt)+ 16h^3 p\sinh (8pt)\right ]
\end{eqnarray}

\begin{eqnarray}
I_4=R_{;a}^{a}&=&\frac{e^{-2ht-\frac{h^2+4p^2}{2}r^2}}{8}h^2{\rm sech}^4(2 p t)
 \left [   -4\,({h^2} - 8\,{p^2})+ (h^2+4p^2)^2r^2 + \right. \nonumber \\
&&\left. \{-4(h^2+ 8 p^2)+(h^2+4p^2)^2r^2   \}{\rm cosh(4pt)}
-8hp\sinh (4pt)\right ]
\end{eqnarray}

%%%%%%%%%%%%%%%%%%%%%%%%%%%%%%%%%%%%%
\section{Petrov      classification}
%%%%%%%%%%%%%%%%%%%%%%%%%%%%%%%%%%%%%%

 Because the solution given here is conformally
related to that of Ref. \cite{mars}, both  have the same Petrov type, namely, of type I except at r=0,
where they  are of  type D in accordance with the theorems  of Ref.\cite{kol} for spaces with symmetry $G_2$. 
The Petrov type can be seen from the non vanishing components  of the Weyl tensor in 
the obvious null tetrad for the solution,
\begin{eqnarray}
\Psi_0 &=& \frac{p^2 e^{(p+s^2/2) r^2}}{\cosh^3(2 pt)}
\left [(\alpha/2+1)\cosh^2(2\,pt) + 
\alpha p r\cosh(2 p t) \sinh(2\,pt)\right. \nonumber \\
&-&\left. 3 -i \{\alpha p r\cosh(2pt)+3\sinh(2pt)  \}  \right ],
\end{eqnarray}

\begin{eqnarray}
\Psi_2 = \frac{p^2 e^{(p+s^2/2) r^2}}{\cosh^3(2 pt)} 
\left [(\alpha/6+1/3)\cosh^2(2\,pt)-1-i\sinh(2pt)  \right ],
\end{eqnarray}
\begin{eqnarray}
\Psi_4 &=& \frac{p^2 e^{(p+s^2/2) r^2}}{\cosh^3(2 pt)}\left [
(\alpha/2+1)\cosh^2(2\,pt)-\alpha r p\cosh(2 p t) \sinh(2\,pt) 
\right. \nonumber \\
 & &\left.-3 - i  \{-\alpha p r \cosh(2pt)+3  \sinh(2pt)\}\right ],
\end{eqnarray}
\noindent where $\alpha= {(2p+s^2)}/{2p^2}$.

%%%%%%%%%%%%%%%%%%%%%%%%%%%%%%%%%%%%%%%%%
\section{Energy momentum tensor}
%%%%%%%%%%%%%%%%%%%%%%%%%%%%%%%%%%%%%%%%%

We also want to look at the energy momentum tensor of the dilaton field.
 From the field equations we can see that the effective energy momentum tensor is
\begin{equation}
T_{a b}^\phi = -\phi_{;a ;b}+\frac{1}{2} \phi g_{a b}.
\end{equation}
For the present solution we have
\begin{eqnarray}
T_{tt}&=&h [ h +p \tanh(2pt) ],\\
T_{tr}&=& h s r,\\
T_{rr}&=&hp \tanh(2pt) ,\\
T_{\varphi \varphi}&=&\frac{[ \cosh^2(2pt)-p^2r^2 ]
hpr^2\sinh(2pt)}{e^{sr^2}\cosh^3(2pt)},\\
T_{\varphi z}&=&\frac{-hp^2r^2 \sinh(2pt)}{e^{sr^2}\cosh^3(2pt)},\\
T_{z z}&=&\frac{-hp \sinh(2pt)}{e^{sr^2}\cosh^3(2pt)}.
\end{eqnarray}
We can see now that the energy momentum tensor is well behaved without 
any singularity.
%T{}=

If we choose a velocity vector that is proportional to $\phi_{;\mu}$ we can calculate the kinematic quantities as acceleration,rotation and expansion. This last one is

$$
\theta = v^\mu_{;\mu}=  \frac{3h   \cosh(2pt)+ 2p \sinh(2pt)}
{ e^{\frac{ht+s r^2}{2}\cosh^{3/2}(2pt) }
}
$$
We notice  that the expansion has a maximum around $t=0$ and is negligible 
for earlier or later times, more precisely,

\begin{equation}
 \left| \theta e^{\frac{s r^2}{2}} \right| \le  \frac{3|h|+2|p|}{2}
\;{  \rm if   } \;\; |t| \ge {\rm max} \{ \frac{1}{|p|},\frac{2}{|h|}\}.  
\end{equation}
This means that only for a short time this solution has an inflationary period.

Finally we give the solution in the conformally related Einstein frame (E-frame) defined by the new 
metric $g_{\mu\nu}^E$ given by $ g_{\mu\nu}^E = e^{-\phi} g_{\mu\nu}^S$
the solution is 
\begin{eqnarray}
ds^2 &=& e^{s r^2} \cosh(2pt) \left ( -dt^2 + dr^2 \right ) 
+ r^2 \cosh(2pt)
d\varphi^2 + \frac{\left ( dz + p r^2 d\varphi \right )^2 }{\cosh(2pt)} 
,\nonumber \\
\phi&=& {h t},
\end{eqnarray}
This is the solution of Ref.\cite{mars} for general relativity with a stiff fluid.

%%%%%%%%%%%%%%%%%%%%%%%%%%%%%%%%%%%%%%%%%%%%%%%%%%%%%%%%%%%
\section{ Acknowledgment}
%%%%%%%%%%%%%%%%%%%%%%%%%%%%%%%%%%%%%%%%%%%%%%%%%%%%%%%%%%%

I am very grateful to the anonymous referee that made suggestions to improve the present work. This work was partially supported by a CONACYT grant.


\newpage
\begin{thebibliography}{000}



\bibitem{wain}J. Wainwright and S. W. Goode, {Phys. Rev. } {\bf D22}, 1906
(1980).
\bibitem{fei}A. Feinstein and J. M. M. Senovilla, { Class. Quantum Grav.} {\bf 6},
 L89 (1989).

\bibitem{van}N. Van den Bergh and J Skea, { Class. Quantum Grav.} {\bf 9},
 527 (1992).


\bibitem{S1} J.\,M.\,M. Senovilla , { Phys. Rev. Lett.} {\bf 64}, 2219 (1990).
\bibitem{mars} M. Mars, { Phys. Rev. } {\bf D51}, 3989 (1995).


\bibitem{Sa}F. Chinea \, J., L. Fern\'{a}ndez-Jambrina,
J. \,M.\,M., Senovilla,  { Phys. Rev. } {\bf D45}, 481 (1992).

\bibitem{RS}E. Ruiz, J. M. M. Senovilla, { Phys. Rev. } {\bf D45}, 1995 (1992).


\bibitem{W1}J. Wainwright, { J. Phys. A: Math. Gen.} {\bf 14}, 1131 (1981).

\bibitem{lete}P. S. Letelier, { J. Math. Phys.} {\bf 20}, 2078 (1979).
\bibitem{kol} C. Kolassis, Class. Quantum Grav. {\bf 6}, 683 (1989).

\bibitem{gmv1}M. Gasperini, J. Maharana and G. Veneziano, Phys. Lett. {\bf B296}, 51 (1992). 


\bibitem{gmv2}M. Gasperini, J. Maharana and G. Veneziano, Phys. Lett. {\bf B272}, 277 (1997). 






\bibitem{V} G. Veneziano, Phys. Lett. {\bf B406}, 297 (1997).
A. Buonanno, K. A.   Meissner, C. Ungarelli and G. Veneziano, 
CERN-TH/97-124.

\bibitem{B&K} J. Barrow and K. Kunze, Phys. Rev. {\bf D55}, 623 (1997);
J. Barrow and K. Kunze, Phys. Rev. {\bf D56}, 741 (1997).

\bibitem{fein}A. Feinstein, R. Lazkoz and M. A. Vazquez-Mozo, Phys. Rev. {\bf D56}, 5166 (1997).hep-th/9704173.

\bibitem{giovannini} M. Giovannini Regular Cosmological examples of the 
Tree-Level Dilaton-Driven Models, Phys. Rev. {\bf D57}, 7223 (1997). hep-th/9712122.

\bibitem{gas}M. Gasperini, M. Maggiore and  G. Veneziano, Nucl. Phys. {\bf B499}, 315 (1997).CERN-TH/96-267. hep-th/9611039.

\bibitem{kalo}N. Kaloper, R. Madden and K. A. Olive, Nucl. Phys. {\bf B452},677(1995).

\bibitem{eas} R. Easther and K. Maeda, Phys. Rev. {\bf D54}, 7252 (1996).hep-th/9605173. 

\bibitem{action}C. Lovelace, Phys. Lett. {\bf B135}, 75 (1984);
E. S. Fradkin and A. A Tseytlin, Nucl. Phys. {\bf B261}, 1 (1985);
C. G. Callan et al., Nucl. Phys. {\bf B262}, 593 (1985);
A. Sen, Phys. Rev. Lett. {\bf 55}, 1846 (1985).





\bibitem{KK}D. Kramer, H. Stephani, E. Herlt, M. MacCallum, ,
{\it Exact solutions of Einstein's field equations } (Cambridge Univ. Press,
Cambridge, England, 1980).
\bibitem{MCK} C. B. G. McIntosh, Phys. Lett.{\bf A43},33 (1973).
\bibitem{nbd}L. O. Pimentel, Modern Physics Letters  {\bf A12 }, 1865 (1997). 




\end{thebibliography}
\end{document}